\documentstyle[aps,prl,psfig,epsf,preprint]{revtex}

\newcommand{\km}{{k_*}} 

\begin{document}

\title{Pattern formation and localization \\ 
in the forced-damped FPU lattice}

\author{Ramaz Khomeriki$^1$\thanks{Permanent address: Department of Physics, 
Tbilisi State University, Chavchavadze ave. 3, Tbilisi 380028, Georgia;
e-mail: khomeriki@hotmail.com},
Stefano Lepri$^{1,2}$\thanks{e-mail: lepri@avanzi.de.unifi.it},
Stefano Ruffo$^{1,2,3}$\thanks{e-mail: ruffo@avanzi.de.unifi.it}}

\address{$^1$Dipartimento di Energetica ``Sergio Stecco" \\
Universit\'a di Firenze, Via S. Marta, 3 - I-50139 Firenze, Italy}

\address{$^2$ Istituto Nazionale di Fisica della Materia, Unit\`a di Firenze}

\address{$^3$ Istituto Nazionale di Fisica Nucleare, Sezione di Firenze}

\date{\today}

\maketitle

\begin{abstract}

\noindent We study spatial pattern formation and energy localization in the 
dynamics of an anharmonic chain with quadratic and quartic
intersite potential subject to an optical, sinusoidally oscillating field and a
weak damping. The zone-boundary mode is stable and locked to the driving field
below a critical forcing that we determine analytically using an 
approximate model which describes mode interactions. 
Above such a forcing, a standing modulated wave forms for driving frequencies
below the band-edge, while a ``multibreather'' state develops at higher 
frequencies. Of the former, we give an explicit approximate analytical 
expression which compares well with numerical data. At higher forcing 
space-time chaotic patterns are observed. 
\vspace{0.5cm}


\end{abstract}

\section{Introduction}
\label{s:introduction}

The dynamics of classical anharmonic lattices displays a rich variety of features.
Already the simplest models, as the one-dimensional chain of equal-mass oscillators
with nearest-neighbors nonlinear forces, exhibit nontrivial solutions like
anharmonic waves~\cite{kosevich,poggi}, discrete solitons~\cite{flytzanis} and
breathers~\cite{flach}.  Due to its simplicity, one of the most widely studied
examples in this class is the celebrated Fermi-Pasta-Ulam (FPU) model~\cite{fermi},
where the interparticle  potential is a simple fourth-order polynomial in the
relative displacements. Many investigations have focused on the process of energy
equipartition among phonons, after having fed the energy into long wavelength
modes, whose instability leads to the generation of solitons~\cite{lichtenberg}.
More recently the complementary case where the energy is placed into the highest
frequency mode has been considered~\cite{burlakov1,ruffo1,corso}. Above a certain
energy threshold, which vanishes as $1/N$, $N$ being the number of oscillators,
this mode becomes modulationally unstable~\cite{peyrard,page1}, leading to the
growth of spatial modulations of the displacement field with a given finite
wavelength.  The subsequent evolution consists into the creation of localized
structures (envelope solitons) which interact inelastically and coalesce in a few,
large amplitude breathers~\cite{lepri}. These have however a finite lifetime and
decay slowly until the asymptotic state of energy equipartition is attained.

Up to now, processes of this kind have been studied mainly for Hamiltonian
lattices, and a quite natural question is to ask how such phenomena are affected by
both forcing and/or dissipative mechanisms that may arise from the presence of an
external field and of the interaction with other degrees of freedom, respectively.
These issues were addressed by R\"ossler and Page~\cite{page2}, who found that
localized modes can indeed exist in a sinusoidally driven, undamped FPU chain. More
recently, the same authors performed detailed studies of the optical generation  of
such excitations under the action of suitable impulsive  fields~\cite{page3}. Later
on, investigations of models of driven-damped  antiferromagnetic chains~\cite{lai}
opened also the way to some experiments, whose outcomes are interpreted as
manifestations of intrinsic  localization~\cite{schwartz}. Theoretical studies of
the parametrically driven, discrete, nonlinear  Schr\"odinger equation~\cite{hennig}
and of coupled oscillator systems~\cite{floria} have also been recently undertaken
and led to the discovery of periodic, quasiperiodic and even chaotic breathers.

However, the relation between the phenomena observed in the Hamiltonian case
with those appearing for forced-damped lattices has not yet been studied in
detail. Indeed, we expect significative differences due to the creation of
stationary states with nontrivial spatial structures, i.e. pattern formation 
~\cite{lvov,manneville}. In our context, having already studied the process
of formation of stable localized  structures arising from modulational
instability in the conservative case~\cite{ruffo1}, we are strongly motivated
to see how the presence of forcing and damping affects this process. To
remain close to the  Hamiltonian case, we restrict ourselves to the case of
small damping. 

Various types of forcing are in principle possible, depending on the physical
situation under study. However, a general requirement for localization is to excite
band-edge modes. For Klein-Gordon lattices  this is naturally realized using a
spatially uniform driving field, which has been shown to induce interesting pattern
formation phenomena~\cite{burlakov2}. On the other hand, this forcing would be
uneffective for FPU lattices, because, due to the   symmetry of the Hamiltonian, the
zero mode is decoupled (see below). Alternatively, since spatial localization
appears from the instability of band-edge modes, we choose in this paper to drive
the system at the zone boundary wavelength. Moreover, we consider the simplest case
in which the driving field oscillates sinusoidally in time.

In Section II we introduce the model and further comment on some of its features.
After introducing a specific mode expansion of the displacement field, we obtain
approximate equations of motion for such modes in the weak damping limit. Some
details of the numerical simulations are presented as well as the quantities used
to detect localization and to study spectral properties. 

Section III deals with the full characterization of the  weak forcing solution
where the zone boundary mode is locked to the driving field. By increasing the
forcing parameter such a solution becomes unstable. Approximate values of the
critical forcing and frequency  are analytically computed and successfully compared
with numerical results. 

Just beyond this instability more complex spatial patterns are attained by the
system. For small enough frequencies a stable standing wave arises, which we
describe analytically by solving for the fixed point of a suitable truncated mode
expansion (Section IV). When the driving  frequency exceeds a given resonant value
a multibreather, spatially aperiodic  state is instead observed in numerical
simulations (Section V).

Section VI is devoted to some conclusions and to a brief discussion of the
transition to chaos, which is observed by increasing the forcing.

\section{The forced and damped FPU model}
\label{s:FPU}

The equations of motion of forced-damped FPU oscillator chain are
\begin{equation}
\ddot u_n=u_{n+1}+u_{n-1}-2u_n+(u_{n+1}-u_n)^3
+(u_{n-1}-u_n)^3-\gamma
\dot u_n+f\cos (\omega t-\pi n),
\label{fpu}
\end{equation}
where $u_{n}$ denotes the displacement of $n$-th oscillator with respect to its
equilibrium position. Periodic boundary conditions,  $u_{n+N}=u_n$, are
assumed, with $N$ being the number of oscillators.  Dimensionless units are
used such that the masses, the linear and nonlinear force constants and the 
lattice spacing are taken equal to unity. The forcing and damping strengths 
are gauged by the parameters $f$ and $\gamma$, respectively, and $\omega$
is the driving frequency. As already mentioned in the Introduction,
the choice of the forcing with the shortest wavelength is expected to favour 
the growth of localized excitations. 
Physically,  the last term in Eq.~(\ref{fpu}) models the interaction of a uniform 
electric field applied to a chain of alternating opposite charges; it can 
in fact be written as $f (-1)^n\cos\omega t$)~\cite{page2}. Let us stress again
that the more widely studied case of uniform forcing~\cite{burlakov2} is 
not viable for models like~(\ref{fpu}) , because the zero mode is completely 
decoupled from the others as a consequence of the invariance of (\ref{fpu}) 
under the transformation $u_n \rightarrow u_n+const.$

In order to describe the forced oscillations of the system it is convenient
to represent the displacement field in the form
\begin{equation}
u_n=\frac{1}{2}\sum\limits_k\left[a_k \, e^{i(\omega t+kn)} +
a_{-k}^+ \,e^{-i(\omega t-kn)}
\right],
\label{aq}
\end{equation}
where $a_k$ are complex mode amplitudes and $-\pi < k \leq \pi$
the corresponding wavenumber. Throughout the paper we will mainly focus
our attention on the dynamics of the zone-boundary mode, which we refer to
as $\pi$-mode for the sake of brevity.

The equations of motion for the amplitudes $a_k$'s are obtained by substituting
Eq.~(\ref{aq}) into Eq.~(\ref{fpu}). Similarly to what is done for
the undamped case~\cite{flach}, a considerable simplification is achieved
by neglecting higher-order harmonics that are produced by the cubic terms
(the so-called rotating wave approximation). Moreover, in the limit of
weak damping $\gamma \ll \omega$ we find the following set of approximate
equations
\begin{equation}
-2i\omega \dot a_k -i\omega\gamma
a_k=(\omega_k^2-\omega^2)a_k+\delta_{k,\pi}f
+6\sum\limits_{q_1,q_2}G_{q_1,q_2}^k a_{q_1}a_{q_2}a_{q_1+q_2-k}^+,
\label{eqa}
\end{equation}
where $\omega_k^2=2(1-\cos k)$, $\delta_{k,\pi}$ is equal to one for $k=\pi$
and zero otherwise and
$$
G_{q_1,q_2}^k=\frac{1}{4} [1+\cos (q_1+q_2)+\cos (k-q_2)
+\cos (k-q_1)-\cos k-\cos q_1
$$
$$-\cos q_2
-\cos (k-q_1-q_2) ].
$$
Eqs.~(\ref{eqa}) correspond to the positive frequency projection in the base given
in Eq.~(\ref{aq}), the negative frequency ones being obtained by replacing $k
\rightarrow -k$ and taking the complex conjugate.

In the following, we aim at comparing the analytical predictions that can be drawn
from the set of approximate equations (\ref{eqa}) with the direct numerical
simulations of model (\ref{fpu}). We have integrated the equations of motion 
(\ref{fpu}) by means of a fourth-order Runge-Kutta algorithm with a step ranging
between $10^{-2}$ and $10^{-3}$. We have always chosen an initial condition with
all oscillators in their equilibrium position $u_n(0)=0$ and small (of the order of
$10^{-5}$) random Gaussian distributed initial velocities   $\dot u_n(0)$. Several
series of simulations have been performed for different values of the parameters
$\omega$ and $f$ and for fixed $\gamma=0.1$. The latter choice guarantees that the
condition $\gamma\ll \omega$ holds in the resonance regions close to the band-edge,
i.e. $|\omega| \sim \omega_\pi=2$ which is of main interest here. Furthermore, the
resulting  time-scales are reasonably short to allow a real-time analysis for
chains as long as $N=512$. Nonetheless, some of the results reported below have
been checked also for another series of simulations performed with $\gamma=0.01$.

We have monitored the energy density along the chain
\begin{equation}
h_n = \frac{1}{2}\dot u_n^2 +
\frac{1}{2}\left[(u_{n+1}-u_{n})^2+(u_{n}-u_{n-1})^2 \right]+
\frac{1}{4}\left[(u_{n+1}-u_{n})^4+(u_{n}-u_{n-1})^4 \right]~ ,
\end{equation}
as well as the spectrum of mode energies
\begin{equation}
\varepsilon_k \;=\; |\dot U_k|^2 + \omega^2 | U_k|^2
\end{equation}
where
\begin{equation}
U_k = {1\over\sqrt{N}}\sum_{n=1}^N u_n \, e^{ikn}=\frac{\sqrt{N}}{2}
\left( a_k e^{i \omega t}+a_{-k}^+ e^{-i \omega t}\right)
\end{equation}
is the amplitude of the $k$-th Fourier mode, which can be efficiently computed
using a standard FFT routine. In the limit in which the $a_k$'s are
slowly varying with respect to the forcing or reach a stationary value, 
one finds that $\varepsilon_k=N\omega^2|a_k|^2$ is constant in time.
On the other hand, when monitoring quantities like
$h_n$ and/or $u_n$ it is  convenient to observe the 
system at time intervals that are integer multiples 
of the driving period.

\section{Modulational instability of the $\pi$-mode}

Let us begin  by considering solutions where only
the $\pi$ mode is excited. Such solutions are numerically observed to exist
and be stable below a certain critical forcing $f_{cr}$  that depends  on both
$\omega$ and $\gamma$. Indeed, in the typical simulation described at the end of
the previous Section, all modes with $k\ne\pi$ damp out on a time-scale set
by the value of $\gamma$ while $|a_\pi|$ rapidly grows and finally 
appoaches a constant value. Such asymptotic amplitude $a_{\pi}$ is obtained 
by solving for the stationary solution of Eqs.~(\ref{eqa}), which corresponds 
to an oscillation locked  to the driving field with a constant phase lag. We get
\begin{equation}
a_{\pi}=\frac{f}{\omega^2-4-12|a_\pi|^2-i\gamma\omega} \quad,
\label{5}
\end{equation}
and writing $a_\pi=|a_\pi|\exp(i \theta_\pi)$ one gets
\begin{equation}
\theta_\pi={\rm atan} \left( \frac{\gamma\omega}{\omega^2-4-12|a_\pi|^2} \right)~,
\label{phase}
\end{equation} 
where the squared modulus $z=|a_\pi|^2$ is the solution the cubic equation 
\begin{equation}
144 z^3 -24 (4-\omega^2) z^2 +
\left[(4-\omega^2)^2+\gamma^2\omega^2\right] z = f^2 \quad.
\label{cub}
\end{equation}
One can easily ascertain that the latter admits a single real root
only for $|\omega|<\omega_*$ where $\omega_*\simeq 2+\sqrt{3}\gamma/2$, while
three distinct roots may otherwise exist (see the discussion in the following).

Before treating in detail the differences between the two cases, let  us
address the question of stability. This is accomplished by solving the
set of equations obtained linearizing Eqs.~(\ref{eqa}) around the  
stationary solutions (i.e. neglecting all interaction terms which are 
nonlinear in the $a_k$s)
\begin{equation}
-2i\omega \dot a_k-i\omega\gamma a_k=(\tilde\omega_k^2-\omega^2)a_k
+3\omega_k^2 a_\pi^2 \, a_{-k}^+ \quad ,
\label{6}
\end{equation}
where
\begin{equation}
\tilde\omega_k^2 =(1+6|a_\pi|^2) \omega_k^2
\label{71}
\end{equation}
is the frequency of the $k$-th mode shifted by the interaction
with the $\pi$-mode. As usual, Eq. (\ref{6}), together its complex conjugate,
are solved looking for solutions of the form $\exp(\nu_k t)$.
The relevant branch of the eigenvalue spectrum reads
\begin{equation}
\nu_k \; = \;
-{\gamma \over 2} + \frac{1}{2|\omega|}\sqrt{9 \omega_k^4|a_\pi|^4 -
(\tilde\omega_k^2-\omega^2)^2}~.
\label{nuk}
\end{equation}
The growth rate $Re\{\nu_k\}$ is maximal when the
square root in the above expression attains its maximum value, i.e. 
when the resonance condition $\omega= \tilde \omega_\km $ 
holds~\cite{firstnote}. The latter, together with definition (\ref{71}), 
fixes the value of the wavenumber $\km$ of the most unstable mode as
\begin{equation}
\cos \km\; = \; 1-\frac{\omega^2}{2(1+6|a_\pi|^2)} \quad.
\label{km}
\end{equation}

In the case in which a single real root of (\ref{cub})) exists
(i.e. $|\omega|<\omega_*$), the threshold for modulational
instability can be computed explicitly. Indeed, by letting $\nu_k=0$ in 
Eq.~(\ref{nuk}) and using formula (\ref{km}), one gets
\begin{equation}
|a_\pi|_{cr}^2=\frac{\gamma}{3(\omega -2\gamma)}~.
\end{equation}
Finally, from formula (\ref{5}) one derives the value of the critical forcing
\begin{equation}
f_{cr}=\sqrt{\frac{\gamma}{3(\omega-2\gamma)}\left\{ \left[
\omega^2-\frac{4(\omega-\gamma)}{\omega-2\gamma}\right]^2
+\gamma^2\omega^2\right\}} \quad.
\label{fcrit}
\end{equation}

\begin{figure}
\centering\psfig{file=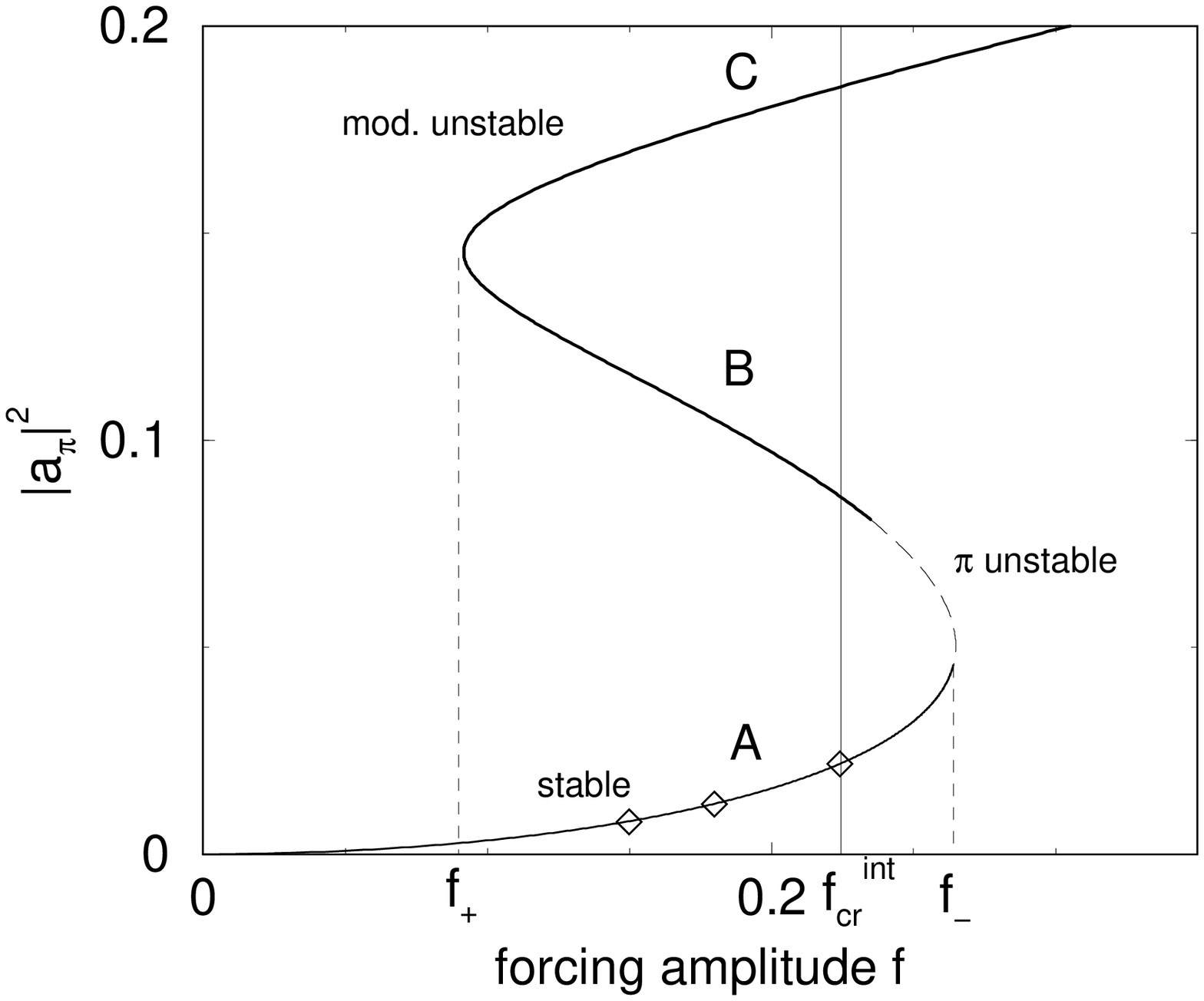,width=8cm}
\noindent
\caption{Squared amplitude of the multiple fixed-point solutions
vs. the forcing $f$ for $\omega=2.4$ and $\gamma=0.1$.
The solid vertical line represents the critical value of the
forcing $f_{cr}^{int}=0.224$. Dashed lines are the boundaries for the
range $f_+<f<f_-$ where three solutions exist. The letters A,B,C
denote the three solutions from the smallest to the highest
amplitude one.}
\label{cubic}
\end{figure}

In the case where three solutions exist ($|\omega|>\omega_*$), the
stability properties must be considered separately for each of them.
Let us discuss this issue with reference to the case $\omega=2.4$,
illustrated by the graph in Fig.~\ref{cubic}.
Here the three solutions, which we label A, B and C, coexist in
the range $f_+<f<f_-$. The values of $f_\pm$ can be
computed from Eq.~(\ref{cub}), since they correspond to the amplitudes
\begin{equation}
|a_{\pi}^{\pm}|^2=\frac{2(\omega^2-4)\pm \sqrt{(\omega^2-4)^2
-3 \gamma^2 \omega^2}}{36}~.
\end{equation}
Hence, from formula (\ref{5}) we obtain
\begin{equation}
f_\pm=|a_{\pi}^{\pm}|\sqrt{(\omega^2-4-12|a_{\pi}^{\pm}|^2)^2+
\gamma^2 \omega^2}~.
\label{bounds}
\end{equation}

Looking at the spectrum (\ref{nuk}), it turns out that the largest amplitude
solution C is always modulationally unstable. The intermediate amplitude solution B
is modulationally unstable above $|a_\pi|^2=(\omega^2-4)/24$ (corresponding to
$f=0.235$ in Fig.~\ref{cubic}) while it has the maximal growth  rate at $\pm\pi$
below this point. The smaller amplitude solution A is instead found to be always
stable. Since we choose to run the dynamics starting always with almost zero mode
amplitudes, it would be natural to conclude that the resulting trajectory converges
to the solution A up to its existence boundary and hence that $f_{cr}=f_-$. The
numerical simulations show that this conclusion is not actually correct. Indeed,
the system approaches the stable solution A only up to a critical value of the
forcing $f_{cr}^{int}$ which is definitely smaller than $f_-$ (see the diamonds in
Fig.~\ref{cubic}). Beyond such a value, the $\pi$-mode undergoes modulational
instability similar to the previous case.

An interpretation of this phenomenon can be given in the following
way. Let us consider the equations of motion for the ``internal'' dynamics
of the $\pi$-mode that can be derived from Eq.~(\ref{eqa}),
\begin{equation}
-2i\omega\dot{a_\pi}-i\omega\gamma a_\pi=
(4-\omega^2)a_\pi+f+12a_\pi|a_\pi|^2~.
\label{internal}
\end{equation}
This nonlinear equation provides a good approximation of the dynamics as long as 
all the other modes are not significantly excited. Its numerical solution
indicates  that for $f=f_{cr}^{int}(\omega)$ the initial condition $a_\pi(0)=0,
\dot a_\pi(0)=0$  exits the basin of attraction of the fixed point corresponding to
the solution A. Such a value corresponds pretty well to the actual critical forcing
numerically determined for the full system (see the solid vertical line at
$f_{cr}^{int}$ in Fig.~\ref{cubic}). In other words, at $f_{cr}^{int}$ the dynamics
leaves the lowest amplitude solution A because of this internal instability and
``jumps'' into the modulationally unstable region. Since a fixed point will not be
subsequently approached, we do not expect that the corresponding spectrum of growth
rates will be described by formula (\ref{nuk}). Nonetheless, a reasonable
qualitative agreement between the two is observed, being both characterized by a
sharp maximum around the most unstable mode with some broadened band around it.

Finally, the results described above are summarized in Fig.~\ref{stability}, where
we show the control parameter space $(f,\omega)$. The dashed line  for $|\omega|<
\omega_*$ is the theoretical expression (\ref{fcrit}) for  the modulational
instability valid when only one solution is present. At $\omega_*$ two full lines
start, given by formula (\ref{bounds}), which bound the region where three
solutions occur. Inside this region, the dotted line $f_{cr}^{int}$ is the one
obtained numerically by looking at the internal $\pi$-mode instability, as we have
described above. The numerical data (full triangles) were obtained by looking at
the incipient modulational instability of the full system. An excellent agreement
(within some percent) with the theoretical results  is  observed in the  considered
frequency range. We have also checked that the critical wavenumber $\km$ is
accurately predicted by Eq.~(\ref{km}). We thus conclude that both the theory
developed for the case of one solution and the approximate description of the
internal instability when three solution are present are basically correct.

\section{Standing nonlinear waves} 

\begin{figure}
\centering\psfig{file=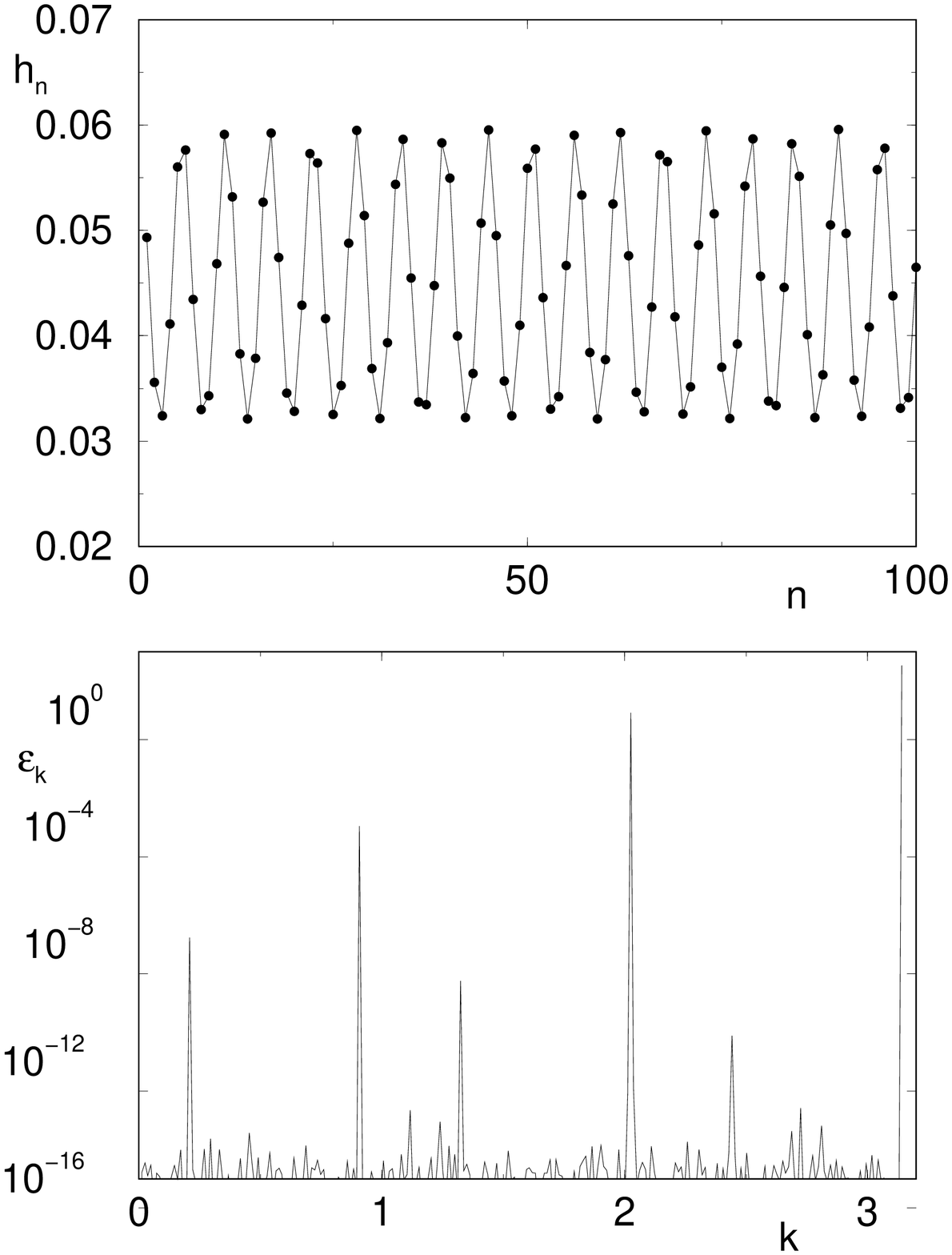,width=10cm}
\noindent
\caption{
Pattern generated after the modulational instability for $\omega=1.8$, 
$f=0.150$ ($f_{cr}=0.148$). This pattern stabilizes
at $t \simeq 10^4$. In the upper panel we show the energy density in a part of
a chain of 512 particles, in the lower the corresponding mode energy spectrum
in  lin-log scale.}
\label{modwave}
\end{figure}

Let us now describe the states forming just above the threshold $f_{cr}$,  after
the development of the modulational instability. Two different behaviors appear
depending whether the driving frequency lies below  or above a resonance frequency,
which for small $\gamma$ is very close to the upper band edge $\omega \simeq 2$. In
this Section we discuss the first case. As expected from the spectrum (\ref{nuk}),
the development of the instability leads to a fast growth of the modes belonging to
the unstable band around $\km$ (and also its harmonics, see below). Afterwards, the
main band shrinks until the system saturates to the asymptotic state. The result is
basically a modulated standing wave locked in time with the forcing field (see the
example illustrated in Fig.~\ref{modwave}) and the wavenumber of the modulation is
indeed very close to the expected value $\pi-\km$ with $\km$ given by
Eq.~(\ref{km}). Furthermore, this state appears to be stable, at least on the time
scale considered in the simulations.  For instance, the wave profile shown in
Fig.~\ref{modwave} remains unaltered up to $2 \, 10^5$ time units, i.e. for more
than $6 \, 10^4$ driving periods.

An approximate theoretical analysis can explain the formation of
this pattern. Indeed, in view of the above results, it is reasonable to look for a
simplified description that neglects all  the modes but the $\pi$-mode and the
most unstable one with wavenumber $\km$. Under such assumptions and taking into
account the resonance condition we obtain from Eqs.~(\ref{eqa}) the following 
coupled equations for $a_\pi$ and $a_\km$:
\begin{eqnarray}
&& -2i\omega\dot{a_\pi}-i\omega\gamma a_\pi =(4-\omega^2)a_\pi +f
+12a_\pi|a_\pi|^2 +6\omega_{\km}^2a_\pi^+a_{\km}a_{-\km} +
12\omega_{\km}^2a_\pi|a_{\km}|^2. \nonumber\\
&& -2i\omega \dot a_\km-i\omega\gamma a_\km =
3 \omega_\km^2 a_\pi^2  a_{-\km}^+
+\frac{9}{4}\omega_{\km}^4  a_{\km}|a_{\km}|^2 \quad.
\label{add1} 
\end{eqnarray}
The first equation is nothing but the modified version of Eq.~(\ref{internal}) 
with the interaction terms between $\pi$ and $\km$ modes included. 
The stationary solutions are determined letting $a_\km=|a_\km|\exp(i
\theta_\km)$ and first solving the second of Eqs.~(\ref{add1}) 
\begin{equation}
|a_{\pm\km}|^2 \; = \; \frac{4}
{3\omega_\km^2}\sqrt{|a_\pi|^4-|a_\pi|_{cr}^4}~, \qquad
\sin 2(\theta_\pi -\theta_{\pm \km}) = - \frac{|a_\pi|_{cr}^2}{|a_\pi|^2}\quad.
\label{10}
\end{equation}
Substituting the latter in the first of Eqs.~(\ref{add1}) we get the 
stationary value of $|a_\pi|$ above threshold 
$$
|a_\pi|^2\Biggl[\left(\omega^2-4-12|a_\pi|^2-16\sqrt{|a_\pi|^4-|a_\pi|_{cr}^4}+8
\frac{|a_\pi|^4-|a_\pi|_{cr}^4}{|a_\pi|^2}\right)^2+
$$
\begin{equation}
\left(\gamma\omega+8\sqrt{|a_\pi|^4-|a_\pi|_
{cr}^4}\right)^2\Biggr]=f^2, 
\label{add2}
\end{equation}
Solving this equation, we can thus get $|a_{\pm\km}|$ and 
$\theta_{\pm\km}$ from Eq.~(\ref{10}). A fourth equation, which we do not 
explicitly display here for the sake of brevity, allows to compute $\theta_\pi$ 
as well.
 
The stationary solution obtained above corresponds to a displacement 
field which can be derived using expansion (\ref{aq}) 
\begin{equation}
u_n \;=\; |a_\pi| (-1)^n \cos(\omega t+\theta_\pi)  + 2 |a_\km| 
\cos(\km n)\cos(\omega t+\theta_\km) \quad,
\label{wave}
\end{equation}
which is precisely a standing modulated wave. These analytical results
are also in reasonable quantitative agreement with numerical
data. For example, in the case of Fig.~\ref{modwave} 
($\omega=1.8$, $f=0.150$ and $\gamma=0.1$) we get
\begin{eqnarray}
&&
|a_ \pi|^2  = 0.02085  \qquad |a_{\km}|^2= 4.88 \, 10^{-4} \quad {\rm
(numerical)} \\
&&
|a_ \pi|^2  = 0.02085  \qquad |a_{\km}|^2= 4.27 \, 10^{-4} \quad {\rm
(theoretical)}~, 
\label{data} 
\end{eqnarray}
the relative deviations being as expected of the order $\gamma/\omega$.

The above expressions are simplified close to threshold and sufficiently far from
resonance. Indeed in this case $\theta_\pi$ is vanishingly small
(see Eq.~(\ref{phase})) and therefore one gets, from formulas (\ref{10}), 
$\theta_\km \simeq \pi/4$ and the approximate solution
\begin{equation}
u_n \;\simeq\; |a_\pi| (-1)^n \cos(\omega t)  + 2 |a_\km| 
\cos(\km n)\cos(\omega t+\frac{\pi}{4}) \quad. 
\end{equation}
This solution compares well with numerical
data. Looking at the pattern at times which are integer multiples 
of $2\pi/\omega$, again for the example of Fig.~\ref{modwave}
with the values in formula (\ref{data}). 
\begin{eqnarray}
&& u_n \simeq 0.1437 (-1)^n + 0.0292 \cos(2.05 n) \quad {\rm (theoretical)}
\nonumber\\  
&& u_n = 0.142 (-1)^n + 0.0359 \cos(2.025 n) \quad {\rm (numerical)}~,
\end{eqnarray}
which shows a good agreement.

Before concluding this Section, let us discuss the issue of higher-order 
corrections to the solution (\ref{wave}). Indeed,  besides the main component at
$\km$, the nonlinear terms induce the presence of several (exponentially small)
harmonics whose wavenumber can be expressed as (recall that $-\pi< k \le \pi$)
\begin{equation}
k_n=\km n+(n-1)\pi, \qquad n=2,~3....
\label{tt}
\end{equation}
Their amplitudes can be computed perturbatively
from the stationary solution of Eq.~(\ref{eqa}). For instance, the first
harmonic ($n=2$) is evaluated as a function $a_\pi$ and $a_\km$.
We give here for completeness the explicit expressions 
of the first and second harmonics.
\begin{eqnarray}
|a_{k_2}| &=&\frac{18\cos \km(\cos \km-1)}
{\omega_{k_2}^2-\omega^2}|a_\pi||a_{\km}|^2, \\
\qquad
|a_{k_3}|&=&\frac{3|1-6\cos ^2\km+2\cos ^3\km|}
{|\omega_{k_3}^2-\omega^2|}|a_{\km}|^3.
\label{pp}
\end{eqnarray}
Higher-order harmonics ($n>2$) can be thus computed recursively in a similar
way obtaining the general result
\begin{equation}
a_{k_n}=\frac{6}
{\omega_{k_n}^2-\omega^2}\sum\limits_{q_n,p_n,r_n<k_n}G_{q_n,p_n}^{k_n}
a_{q_n}a_{p_n}a_{-r_n}^+, \qquad q_n+p_n+r_n=k_n.
\label{rr}
\end{equation}
Notice that an infinite number of harmonics is expected here. This shows that
the waves we are dealing with are more general than those previously
reported in the literature for damped-driven Klein-Gordon lattices
\cite{burlakov2}. In this latter case, the modulation has a finite number of
harmonics due to the peculiar mutual relationship among modes $0$, $\pi/2$ and
$\pi$ that allows for solution where no other modes are excited.

\section{Multibreather states}

\begin{figure}
\centering\psfig{file=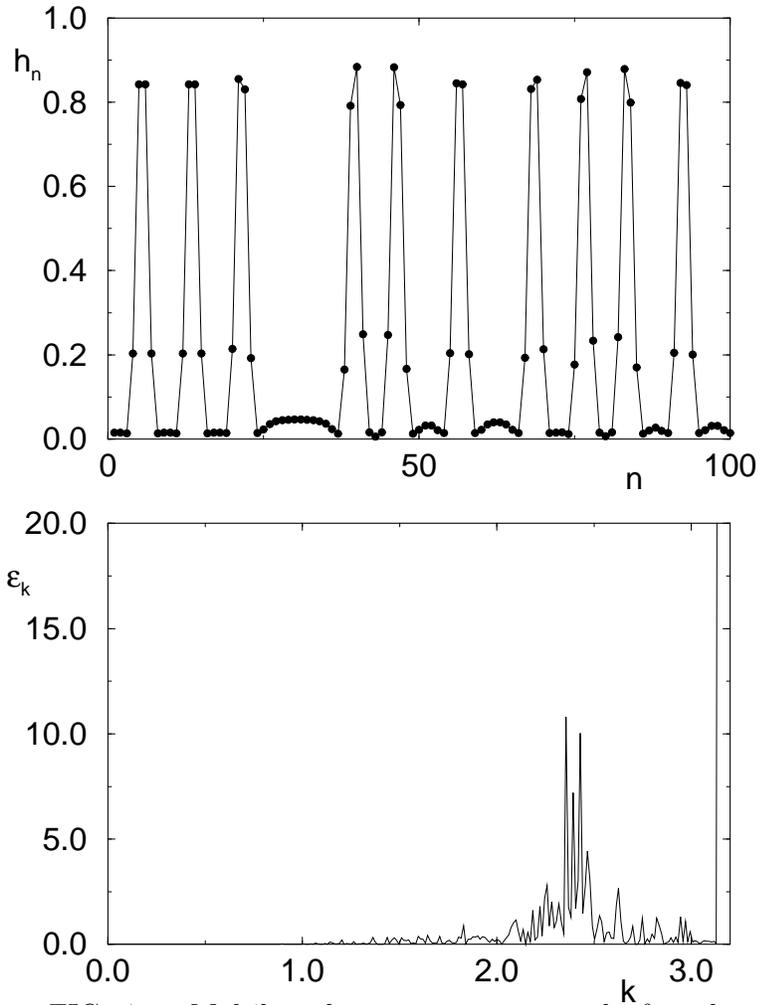,width=10cm}
\noindent
\caption{
Multibreather pattern generated after the modulational instability for  
$\omega=2.4$, $f=0.2250$ ($f_{cr}=0.2245$). This pattern stabilizes
at $t \simeq 5\, 10^3$. In the upper panel we report the energy density, in the
lower the mode energy spectrum.}
\label{multibre}
\end{figure}

A different scenario is observed for driving frequency above resonance, which we
briefly describe here with reference to the case $\omega=2.4$, illustrated in
Fig.~\ref{multibre}.  The instability described at end of Section III produces, on
a relatively short time scale ($t\sim 10^2$), a disordered assembly of sharply
localized structures in a similar way  to what is observed for the  undriven
case~\cite{ruffo1,lepri}. On longer time  scales ($t\sim 10^3$), a further stage
follows in which the localized peaks  arrange themselves until they eventually
reach an asymptotic state (see the upper panel  of Fig.~\ref{multibre}). This is a
sort of ``multibreather'' state, i.e. an array of {\it unevenly spaced} breathers. 
Remarkably, such a complex solution appears to be stable as the previously
illustrated nonlinear wave. For instance, the pattern in  Fig.~\ref{multibre} has
been observed to persist up to a time $2.0\, 10^5$, i.e. more than $7 \, 10^4$
driving periods.

\begin{figure}
\centering\psfig{file=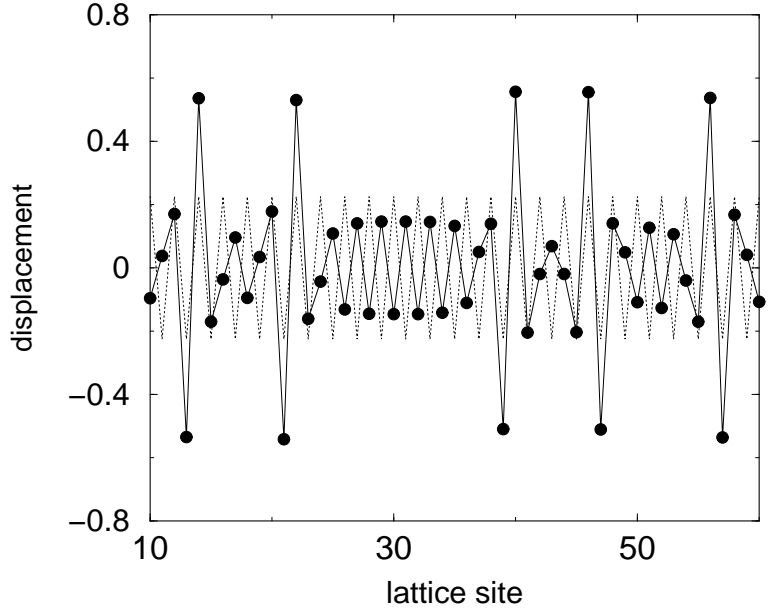,width=10cm}
\noindent
\caption{
Displacement pattern corresponding to the case of  Fig.~\ref{multibre}. The dashed
line is the istantaneous configuration of  the driving field showing relative phase
relations. } 
\label{disp} 
\end{figure}

The corresponding displacement field in Fig.~\ref{disp} reveals that each  breather
is pretty similar to the ``even-parity'' localized modes found in the Hamiltonian
case~\cite{page1},  the main difference being the presence of a $\pi$-mode
background induced by the field. No ``odd-parity'' mode is generated by the above
mechanism in all the examined cases. Actually, a closer inspection  reveals that two
distinct solutions of slightly different amplitude are  present.  Moreover, apart
from a small lag induced by the damping, the background oscillates always out of
phase with respect to the field as it can be ascertained by comparing the dotted
line in Fig.~\ref{disp} with the corresponding displacement pattern. On the
contrary, the breathers are always in-phase with the field. Accordingly, depending
on the site on which they sit, they can have equal or opposite relative phases.
Although localized solutions oscillating out of phase with the field are known to
exist in similar models~\cite{page2}, they are not detected in the present context. 

The inhomogeneous distribution of vibrational energy is distinctly  reflected in
the mode spectrum (see the lower panel of Fig. \ref{multibre}), which displays a
localized structure and a band broadening as a consequence of the  irregular
spacing among the breathers. 

To give a more firm basis to our numerical observations, it is useful to
briefly  point out some conclusions that can be drawn by a suitable continuum
approximation of the FPU model. In analogy with the approach followed for the
hamiltonian case~\cite{lepri}, one can in fact write the displacement field as
$u_n \;=\; (-1)^n Re\{\psi(x,t)\exp(i\omega t)\}$ where $\psi$ is an envelope 
function which is assumed to be slowly varying in space and time on scales of
the order of the interatomic spacing and the driving period, respectively.
A standard calculation leads to the driven-damped nonlinear Schr\"odinger 
equation
\begin{equation}
2i\omega \dot\psi + (4-\omega^2 +i\omega\gamma)\psi + \psi_{xx}
+12 \psi |\psi|^2 \;=\; f \quad,
\label{nls}
\end{equation}
which in the spatially uniform case reduces to Eq.~(\ref{internal}) for the
$\pi$-mode amplitude. Notice however that this type of description makes sense
only when a relatively  narrow packet of modes close  to $\pm\pi$ is excited
and is thus less general than the one based on Eqs.~(\ref{eqa}).  For a
suitable choice of parameters,  Eq.~(\ref{nls}) admits two soliton  solutions
of different amplitudes~\cite{baras} as well as stable multisoliton  complexes
arising from bifurcation of one of them~\cite{baras2}. Thus,  the localized
states of Fig.~\ref{disp} could be related to such solutions,  at least to the
extent in which our lattice model can be approximated by a continuum equation
like ~(\ref{nls}). Although a more quantitative comparison would be desirable,
this is a solid argument in support of the existence of stable multibreather
complexes.

\section{Conclusions and perspectives}

We have confirmed that modulational instability of zone boundary modes is a relevant
mechanism for the generation of nontrivial spatial structures in discrete
anharmonic lattices. Our results for the externally driven case complement 
previous studies on Hamiltonian models and further show that the interplay of
(almost) resonant forcing and damping can stabilize such  structures.  Our
approximate analytical framework has allowed also to derive the stability chart of
the zone-boundary mode. For forcing frequencies below resonance a modulated wave is
formed after instability, whose approximate analytical expression we  have derived
in terms of mode amplitudes. For frequencies above resonance  a ``multibreather"
state arises of which we have given a phenomenological  characterization. There are
other problems that could be attacked within this approach, like  the interesting
issue of destabilization of the modulated wave. This  could also contribute to a
better understanding of localization mechanisms. Furthermore, this points out the
possibility of generating long-living and complex energy distributions in space for
real ionic crystals under  the action of optical fields.

\begin{figure}
\par
\centering{\psfig{file=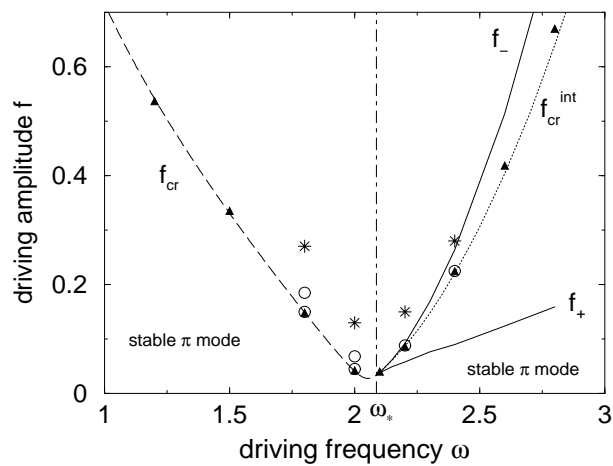,width=8cm}}
\par
\noindent
\caption{
Control parameter plane $(\omega,f)$ for $\gamma=0.1$. The dashed
line for $\omega < \omega_*$ is the critical forcing $f_{cr}$ given by 
formula (\ref{fcrit}) at which modulational instability occurs.
The region for $\omega > \omega_*$ where three solutions exist is bounded by
the solid lines $f_\pm$. The dotted line within corresponds to the
instability threshold $f_{cr}^{int}$ evaluated numerically  from
Eq.~(\ref{internal}). The full triangles are the numerical estimates of
$f_{cr}$. Open circles left (resp. right) of the
$\omega=\omega_*$ line denote  points where standing waves (resp. breathers)
occur. The stars are some parameter values for  which  chaos is
detected.}
\label{stability}
\end{figure}

\begin{figure}
\centering\psfig{file=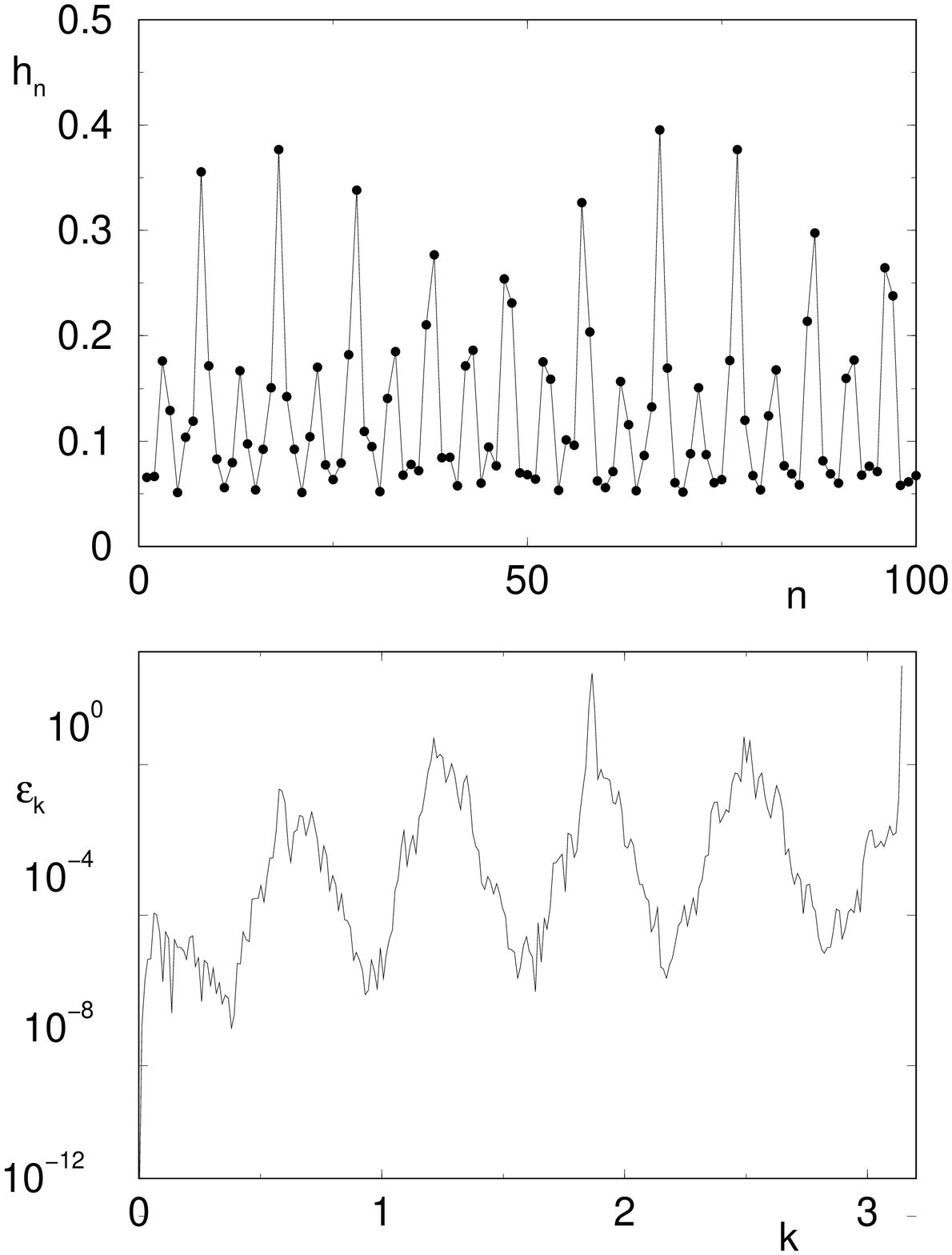,width=10cm}
\noindent
\caption{
Chaotic state obtained  for $\omega=1.8$, $f=0.27$. In the upper panel we show 
a snapshot of the energy density along the chain at $t=5200$ and in
the lower the corresponding mode energy spectrum in lin-log scale.
}
\label{chaotic}
\end{figure}

Let us also briefly comment on the fate of the two typical spatial patterns 
described above. As expected, increasing the forcing leads to their destabilization
and eventually to the transition to chaos (see the stars in Fig.~\ref{stability} as
well as Fig.~\ref{chaotic}). This bears a strong analogy with the {\it
parametrically} driven-damped nonlinear Schr\"odinger equation where stable patterns
precede the onset of chaotic behavior~\cite{baras3}. This is a further indication 
that a considerable insight on the dynamics of the discrete model may be achieved
from Eq.~(\ref{nls}).  Besides of this, it has been suggested on quite general
grounds that systems obeying the symmetry $u_n\to u_n+const.$ should show an abrupt
transition to a soft-turbulent state without displaying any intermediate spatial
pattern~\cite{tribelsky}. This is in apparent contradiction with  the existence of
stable patterns found in our numerics. On the other hand, performing simulations at
$\gamma=0$ the transition to chaos occurs as soon as $f\ne 0$, since the
modulational pattern does not attain any regular asymptotic state. We can therefore
conjecture that the presence of friction alters the nature of the transition in a
way that remains to be understood.

\acknowledgments

We acknowledge useful discussions with the members of the research group {\it
Dynamics of Complex Systems} in Florence as well as a partially support by
the INFM project {\it Equilibrium and nonequilibrium dynamics in condensed
matter}. This work is also part of the EC network LOCNET, Contract No.
HPRN-CT-1999-00163 and of the MURST-COFIN00 project "Chaos and localization in
classical and quantum mechanics".
R. Kh. is obliged to CNR-NATO for the visiting scientist fellowship award
providing the financial support for his stay in Florence University.

\begin{thebibliography}{99}
\bibitem{firstnote}
Notice that this resonance condition is actually approximate. Indeed, upon
setting to zero the derivative of $\nu_k $ with respect to $k$, one finds
it to be  valid up to a small correction (of order $|a_\pi|^4$) which
we neglect for the sake of simplicity in deriving the analytical
solution.
\end{thebibliography}
\end{document}